\documentclass[11pt]{article}
\usepackage{amssymb,graphicx}
\usepackage{tabularx}
\usepackage{array}
\def\squarebox#1{\hbox to #1{\hfill\vbox to #1{\vfill}}}
\newcommand{\qed}{\hspace*{\fill}
            \vbox{\hrule\hbox{\vrule\squarebox{.667em}\vrule}\hrule}\smallskip\newline}
\newtheorem{THEOREM}{Theorem}
\newenvironment{theorem}{\begin{THEOREM} \hspace{-.85em} {\bf :} \rm}                        {\end{THEOREM}}
\newtheorem{LEMMA}{Lemma}
\newenvironment{lemma}{\begin{LEMMA} \hspace{-.85em} {\bf :} \rm}                      {\end{LEMMA}}
\newtheorem{COROLLARY}{Corollary}
\newenvironment{corollary}{\begin{COROLLARY} \hspace{-.85em} {\bf :} \rm}                          {\end{COROLLARY}}
\newenvironment{proof}{\noindent {\bf Proof:}}{\qed}
\newtheorem{DEFINITION}{Definition}

\newtheorem{CLAIM}{Claim}
                      {\end{CLAIM}}

\begin{document}

\title{An $\tilde{O}(n^{2.5})$-Time Algorithm for Online Topological Ordering}

\author{Hsiao-Fei Liu$^1$ and Kun-Mao Chao$^{1,2,3}$
\\\\
$^1$Department of Computer Science and Information Engineering \\
$^2$Graduate Institute of Biomedical Electronics and
Bioinformatics \\
$^3$Graduate Institute of Networking and Multimedia \\
National Taiwan University, Taipei, Taiwan 106\\
}

\maketitle

\begin{abstract}
We present an $\tilde{O}(n^{2.5})$-time algorithm for maintaining
the topological order of a directed acyclic graph with $n$ vertices
while inserting $m$ edges. This is an improvement over the previous
result of $O(n^{2.75})$ by Ajwani, Friedrich, and Meyer.
\\

\end{abstract}

\pagestyle{myheadings}

\thispagestyle{plain}

\markboth{}{Online Topological Ordering}

\section {Introduction}
A topological order $T$ of a directed acyclic graph (DAG) $ G = (V,
E)$ is a linear order of all its vertices such that if $G$ contains
an edge $(u, v)$, then $T(u) < T(v)$. In this paper we study an
online variant of the topological ordering problem in which the
edges of the DAG are given one at a time and we have to update the
order $T$ each time an edge is added. Its practical applications can
be found in \cite{Alpern,Marchetti-Spaccamela 1993,Pearce}. In this
paper, we give an $\tilde{O}(n^{2.5})$-time\footnote{The symbol
$\tilde{O}$ means $O$ with log factors ignored. Depending on the
implementation, the runtime may vary from $O(n^{2.5}\log^2n)$ to
$O(n^{2.5}\log n)$.} algorithm for online topological ordering.

\subsection {Related Work}
Alpern et al. \cite{Alpern} gave an algorithm which takes $O(
||\delta||\log||\delta||)$ time for each edge insertion, where
$||\delta||$ is a measure of the change. (For a formal definition of
$||\delta||$, please see \cite{Alpern,{Pearce}, Ramalingam}.) Pearce
and Kelly \cite{Pearce} proposed a different algorithm which needs
slightly more time to process an edge insertion in the worst case
than the algorithm given by Alpern et al. \cite{Alpern}, but showed
experimentally their algorithm perform well on sparse graphs.

Marchetti-Spaccamela et al. \cite{Marchetti-Spaccamela 1996} gave an
algorithm which takes $O(mn)$ time for inserting $m$ edges. Katriel
\cite{Katriel 2004} showed that the analysis is tight. Katriel and
Bodlaender \cite{Katriel 2006} modified the algorithm proposed by
Alpern et al. \cite{Alpern}, which is referred to as the
Katriel-Bodlaender algorithm. Katriel and Bodlaender proved that
their algorithm has both an $O(\min \{m^{3/2}\log n,m^{3/2}+n^2\log
n\})$ upper bound and an $\Omega(m^{3/2})$ lower bound on runtime
for $m$ edge insertions. Katriel and Bodlaender also analyzed the
complexity of their algorithm on structured graphs. They showed that
the Katriel-Bodlaender algorithm runs in time $O(mk \log^{2} n)$
where $k$ is the treewidth of the underlying undirected graph and
can be implemented to run in $O(n \log n)$ time on trees. In
\cite{Liu}, we proved that the Katriel-Bodlaender algorithm takes
$\Theta(m^{3/2}+ mn^{1/2}\log n)$ time for inserting $m$ edges.
Recently, Ajwani~et~al.~\cite{Ajwani} proposed an $O(n^{2.75})$-time
algorithm, independent of the number of edges $m$ inserted. To the
best of our knowledge, it is the best result for dense DAGs.


\section {Algorithm}
We keep the current topological order as a bijective function
$T:V\rightarrow [1..n]$. Let $d(u,v)$ denote $|T(v)-T(u)|$,
$u\rightarrow v$ express that there is an edge from $u$ to $v$,
$u\rightsquigarrow v$ express that there is a path from $u$ to $v$
and $u<v$ be a short form of $T(u)<T(v)$. Let $n^{0.5}< t_0
<t_{1}<t_{2}<\ldots<t_{p-1}<t_p<t_{p+1}=n$, where $p=O(\log n)$ is a
nonnegative integer. In Section~\ref{discussion}, we shall show how
to determine the values of these parameters.

Figure~\ref{algo} gives the pseudo code of our algorithm. $T$ is
initialized with the topological order of the starting graph.
Whenever an edge $(u,v)$ is inserted into the graph,
\textsc{Insert}$(u,v)$ is called. If $u<v$, then
\textsc{Insert}$(u,v)$ does not change $T$ and simply insert the
edge into the graph. If $u>v$, then \textsc{Insert}$(u,v)$ calls
\textsc{Reorder}$(v,u,0,0)$ to update $T$ such that $T$ is still a
valid topological order and $T(u)<T(v)$. After the call to
\textsc{Reorder}$(v,u,0,0)$, \textsc{Insert}$(u,v)$ can safely
insert the edge into the graph.

It remains to explain how the procedure
\textsc{Reorder}$(u,v,f_1,f_2)$ works. The duty of the procedure
\textsc{Reorder}$(u,v,f_1,f_2)$ is to update $T$ such that $T$ is
still a valid topological order and $T(u)>T(v)$. The flag $f_1=1$
indicates that the set $A' = \{w \mbox{ : } u\rightarrow w \mbox{
and } w\leq v \}$ has been known to be empty.  The flag $f_2=1$
indicates that the set $B' = \{w \mbox{ : } w\rightarrow~v \mbox{
and } w\geq u \}$ has been known to be empty. If $T(u)>T(v)$, then
we directly exit. Otherwise, there are two cases to consider:
\begin{enumerate}

\item[1:] $t_i<d(u,v)\leq t_{i+1}$ for some $i=0,\ldots,p$. In this case, we first have to compute $\hat{A}_i = \{w \mbox{ : } u\rightarrow w
\mbox{, } d(u,w) \leq t_{i} \mbox{, and } w < v\}$ and $\hat{B}_i =
\{w \mbox{ : } w\rightarrow v \mbox{, } d(w,v) \leq t_{i} \mbox {,
and } w > u\}$. If $\hat{A}_i=\emptyset$ and $f_1=0$, then we still
have to compute $\hat{A}_{i+1}= \{w \mbox{ : } u\rightarrow w
\mbox{, } d(u,w) \leq t_{i+1} \mbox{, and } w < v\}$ and set
$A=\hat{A}_{i+1}$; otherwise, we directly set $A=\hat{A}_i$.
Similarly, if $\hat{B}_i=\emptyset$ and $f_2=0$, then we still have
to compute $\hat{B}_{i+1}= \{w \mbox{ : } u\rightarrow w \mbox{, }
d(u,w) \leq t_{i+1} \mbox{, and } w < v\}$ and set
$B=\hat{B}_{i+1}$; otherwise, we directly set $B=\hat{B}_i$.

\item[2:] $d(u,v)\leq t_{0}$. In this case we directly set $A=\hat{A}_0 = \{w
\mbox{ : } u\rightarrow w \mbox{, } d(u,w) \leq t_{0} \mbox{, and }
w < v\}$ and $B=\hat{B}_0 = \{w \mbox{ : } w\rightarrow v \mbox{, }
d(w,v) \leq t_{0} \mbox {, and } w > u\}$.

\end{enumerate}

\noindent If both $A$ and $B$ are empty, then we directly swap $u$
and $v$ and exit the procedure. Otherwise, let $T_{origianl}$ be the
topological order at the start of the execution of the procedure.
For each $u' \in \{u\}\cup A$, considered in order of decreasing
$T_{original}(u')$, we do the following. For each $v'\in\{v': v'\in
B\cup\{v\} \mbox{ and } T_{original}(v')>T_{orginal}(u')\}$,
considered in order of increasing $T_{original}(v')$, recursively
call \textsc{Reorder}$(u',v',f'_1,f'_2)$. The first flag $f_1'$ is
set to 1 if and only if $u'=u$ and $A=\emptyset$, and the second
flag $f_2'$ is set to 1 if and only if $v'=v$ and $B=\emptyset$.

\vspace{1em}
 \noindent\textbf{The idea behind the algorithm.} Our
algorithm broadly follows the algorithm by
Ajwani~et~al.~\cite{Ajwani}. The main difference is that
Ajwani~et~al. always set $A$ to $\hat{A}_{i+1}$ and $B$ to
$\hat{B}_{i+1}$ during the execution of \textsc{Reorder} but we set
$A$ to $\hat{A}_{i+1}$ only if $\hat{A}_i = \emptyset$ and $B$ to
$\hat{B}_{i+1}$ only if $\hat{B}_i = \emptyset$. We shall prove that
the total number of calls to \textsc{Reorder} won't increase
(bounded above by $O(n^2)$) by introducing this modification. Thus
intuitively, our algorithm should run faster because in each call to
\textsc{Reorder} we might only need to compute $\hat{A}_i$ and
$\hat{B}_i$ instead of $\hat{A}_{i+1}$ and $\hat{B}_{i+1}$.

\begin{figure}[!ht]
    \noindent\hrulefill\vspace{-0.6cm}
    \begin{tabbing}
    \hspace*{1em} \= \hspace*{1em} \= \hspace*{1em} \= \hspace*{1em}
    \= \hspace*{1em} \= \hspace*{1em} \kill \\
    \textsc{Insert}$(u,v)$ \\
    \ \>(* Insert edge $(u,v)$ and calculate new topological ordering *)\\
    \ 1\> \textbf{if} $v\leq u$ \textbf{then} \textsc{Reorder}$(v,u,0,0)$\\
    \ 2\>  insert edge $(u,v)$ in graph\\

    \hspace*{1em} \kill \\
    \textsc{Reorder}$(u,v,f_1,f_2)$ \\
    \ \>(* Reorder vertices between $u$ and $v$ such that $v\leq u$ *)\\
    \ 1\> \textbf{if} $v\leq u$ \textbf{then} exit\\
    \ 2\> \textbf{if} $t_{i}<d(u,v)\leq t_{i+1}$\\
    \ 3\>\> \textbf{then}\\
    \ 4\>\>\> $\hat{A}_i := \{w \mbox{ : } u\rightarrow w \mbox{, } d(u,w) \leq  t_{i} \mbox{, and } w < v\}$\\
    \ 5\>\>\> $\hat{B}_i := \{w \mbox{ : } w\rightarrow v \mbox{, } d(w,v) \leq  t_{i} \mbox {, and } w > u\}$\\
    \ 6\>\>\> $A :=\hat{A}_i$ if $A_i\neq\emptyset$ or $f_1=1$; otherwise, $A:=\hat{A}_{i+1}:= \{w \mbox{ : } u\rightarrow w \mbox{, } d(u,w) \leq  t_{i+1} \mbox{, and } w < v\}$\\
    \ 7\>\>\> $B :=\hat{B}_i$ if $B_i\neq\emptyset$ or $f_2=1$; otherwise, $B:=\hat{B}_{i+1}:= \{w \mbox{ : } w\rightarrow v \mbox{, } d(w,v) \leq  t_{i+1} \mbox {, and } w > u\}$\\
    \ 8\>\> \textbf{else}\\
    \ 9\>\>\> $A:=\hat{A}_0:=\{w \mbox{ : } u\rightarrow w \mbox{, } d(u,w) \leq  t_0 \mbox{, and } w < v\}$\\
     10\>\>\> $B:=\hat{B}_0:=\{w \mbox{ : } w\rightarrow v \mbox{, } d(w,v) \leq  t_0 \mbox {, and } w > u\}$\\
     11 \> \textbf{if} $A=\emptyset \mbox{ and } B=\emptyset$\\
     12 \>\> \textbf{then}\\
     13\>\>\> swap $u$ and $v$\\
     14 \>\> \textbf{else}\\
     15\>\>\> \textbf{for} $u'\in \{u\}\cup A$ in decreasing topological order\\
     16\>\>\>\> \textbf{for} $v'\in B\cup\{v\}\wedge v' > u'$ in increasing topological order\\
     17\>\>\>\>\> $f_1':=1$ if ($u=u'$ and $A=\emptyset$); otherwise, $f_1':=0$\\
     18\>\>\>\>\> $f_2':=1$ if ($v=v'$ and $B=\emptyset$); otherwise,  $f_2':=0$\\
     19\>\>\>\>\> \textsc{Reorder}$(u',v', f_1',f_2')$\\
    \end{tabbing}
    \vspace{-0.75 cm}\noindent\hrulefill
    \caption{Our algorithm.}
    \label{algo}
    \end{figure}


\section {Data Structures}

\subsection {Main Data Structures}
In the following, we shall describe the main data structures used in
our algorithm.

The current topological order $T$ and its inverse $T^{-1}$ are
stored as arrays. Thus finding $T(i)$ and $T^{-1}(u)$ can be done in
constant time.

The DAG $G=(V,E)$ is stored as an array of vertices. For each vertex
$u$ we maintain two adjacency lists $InList(u)$ and $OutList(u)$.
The backward adjacency list $InList[u]$ contains all vertices $v$
with $(v,u)\in E$. The forward adjacency list $OutList(u)$ contains
all vertices $v$ with $(u,v)\in E$. Adjacency lists are implemented
by using $n$-bit arrays and support the following operations.

 \begin{enumerate}\itemsep=-4pt
        \item \textsc{List-Insert}: Given a vertex and a list, add the vertex to the list.
        \item \textsc{List-Search}: Given a vertex and a list,
            determine if the vertex is in the list. If yes, return 1. Else, return 0.
 \end{enumerate}
Since the adjacency lists are implemented by using $n$-bit arrays,
it takes $O(1)$ time per \textsc{List-Insert} or
\textsc{List-Search} operation.

\subsection {Auxiliary Data Structures}
In the following we describe some auxiliary data structures which
are used in our algorithm to improve the time complexity. For each
vertex $u$, we maintain two arrays of pails: $InPails(u)[0\cdots
p+1]$ and $OutPails(u)[0\cdots p+1]$. $InPails(u)[i]$ contains all
vertices $v$ with $0 <d(v,u)\leq  t_i$ and $(v,u)\in E$.
$OutPails(u)[i]$ contains all vertices $v$ with $0<d(u,v)\leq t_i$
and $(u,v)\in E$. A vertex $v$ in a pail is stored with its vertex
index (and not $T(v)$) as its key. Pails are implemented by using
balanced binary search trees and support the following operations.

\begin{enumerate}\itemsep=-4pt
        \item \textsc{Pail-Insert}: Given a vertex and a pail, add the vertex to the pail.
        \item \textsc{Pail-Delete}: Given a vertex and a pail, delete the vertex from the pail.
        \item \textsc{Pail-Collect-All}: Given a pail, report all vertices in the pail.
 \end{enumerate}

It takes $O(\log n)$ time per \textsc{Pail-Insert} or
 \textsc{Pail-Delete}
 and $O(1+\gamma)$ time per \textsc{Pail-Collect-All}, where $\gamma$ is the number of vertices in the pail.

\subsection{Instructions for Data Structures} \label{Instructions}
Given a DAG $G$ with a valid topological order and two vertices $u$
and $v$ with $u\not\rightsquigarrow v$, define sorted vertex sets
$\hat{A}_i$ and $\hat{B}_i$, $i=0,\ldots,p+1$, as follows:

 \[
        \begin{array}{ll}
        \hat{A}_i = \{w \mbox{ : } u\rightarrow w \mbox{ and } d(u,w) \leq  t_i \mbox{ and } w < v\} \mbox{ sorted by the
        topological order.}\\
        \hat{B}_i = \{w \mbox{ : } w\rightarrow v \mbox{ and } d(u,w) \leq  t_i \mbox{ and } w > u\} \mbox{ sorted by the
        topological order.}\\
        \end{array}
\]

In the following we discuss how to insert an edge, compute vertex
sets $\hat{A}_i$, and $\hat{B}_i$, and swap two vertices in terms of
the above five basic operations.
\begin{enumerate}\itemsep=-4pt
        \item[a.] Inserting an edge $(u,v)$: This means inserting vertex $v$ to the
            forward adjacency list of $u$ and $u$ to the backward adjacency list of $v$.
            This requires two \textsc{List-Insert} operations and at most
            $2(p+2)$ \textsc{Pail-Insert}
            operations. Thus inserting an edge $(u,v)$ can be done
            in $O(p\log n)=\tilde{O}(1)$ time.
        \item[b.] Computing $\hat{A}_i$ and $\hat{B}_i$: $\hat{A}_i$ can be computed by sorting the vertices in $OutPail(u)[i]$
            and choosing all $w$ with $w<v$. This
            can be done by first calling \textsc{Pail-Collect-All}
            to collect all the vertices in $OutPail(u)[i]$ in $O(t_i)$ time.
            Note that for all these vertices $w$, we have $0<T(w)-T(u)\leq t_i$.
            Thus we can sort these vertices in $O(t_i)$ time by counting sort and then choose all $w$ with $w<v$ in $O(|\hat{A}_i|+1)$ time.
            The total time required to compute $\hat{A}_i$ is $O(t_i+|\hat{A}_i|)=O(t_i)$.
            Similarly, the time required to compute $\hat{B}_i$ is $O(t_i)$.
        \item[c.] Computing $\hat{A}_i$ and $\hat{B}_i$ when $t_i<d(u,v)\leq t_{i+1}$: $\hat{A}_i$ can be computed by sorting the vertices in $OutPail(u)[i]$.
            This can be done by first calling \textsc{Pail-Collect-All}
            to collect all the vertices in $OutPail(u)[i]$ in $O(|\hat{A_i}|+1)$ time, and then sorting these
            vertices in $O((|\hat{A_i}|+1)\log n)$ time.
            Thus the total time required to compute $\hat{A}_i$ is $\tilde{O}(|\hat{A}_i|+1)$.
            Similarly, the time required to compute $\hat{B}_i$ is $\tilde{O}(|\hat{B}_i|+1)$.
        \item[d.] Swapping $u$ and $v$: Without loss
            of generality assume $u<v$.
            When swapping $u$ and $v$, we need to update
            the pails, $T$, and $T^{-1}$. We now show how to update the pails.
            For all vertices $w$ with $T(u)-t_i\leq T(w)<
            \min\{T(u),T(u)-t_i+d(u,v)\}$,
            we delete $w$ from $InPail(u)[i]$ and delete $u$ from $OutPail(w)[i]$.
            For all vertices $w$ with $\max\{T(v)+t_i-d(u,v),T(v)\}< T(w)\leq
            T(v)+t_i\}$,
            we delete $w$ from $OutPail(v)[i]$ and delete $v$ from $InPail(w)[i]$.
            It requires total $O(d(u,v))$
            \textsc{Pail-Delete} operations for each $i$. For all $w$ with $\max\{T(v),T(u)+t_i\}<T(w)\leq T(u)+t_i+d(u,v),$
            if $w$ is in the forward adjacency list of $u$, then insert $w$
            into $OutPail(u)[i]$ and insert $u$ into $InPail(w)[i]$.
            For all $w$ with $T(v)-t_i-d(u,v)\leq T(w)< \min\{T(u),T(v)-t_i+d(u,v)\},$
            if $w$ is in the backward adjacency list of $v$, then insert $w$
            into $InPail(v)[i]$ and insert $v$ into $OutPail(w)[i]$. It
            requires total $O(d(u,v))$ \textsc{List-Search}
            and \textsc{Pail-Insert} operations for each~$i$. In total, we need
            $O(p\cdot d(u,v))$ \textsc{List-Search}, \textsc{Pail-Insert}, and
            \textsc{Pail-Delete} operations. Updating $T$ and
            $T^{-1}$ is trivial and can be done in constant time.
            Thus the total time is $O(p\cdot d(u,v)\log n)=\tilde{O}(d(u,v))$.
 \end{enumerate}

\section {Correctness}
In this section, we shall argue that our algorithm is correct. We
say a call of a recursive procedure leads to an operation ``by
itself'' if and only if this operation is executed during the
execution of this call and not during the execution of subsequent
recursive calls. Given a DAG $G$ with a valid topological order $T$
and two vertices $u, v$ of $G$ with $u<v$, let $A' = \{w \mbox{ : }
u\rightarrow w \mbox{ and } w\leq v \}$ and $B' = \{w \mbox{ : }
w\rightarrow~v \mbox{ and } w\geq u \}$. We say the flag $f_1$ of
the call to \textsc{Reorder}$(u,v,f_1,f_2)$ is correctly set only if
$(f_1\Rightarrow (A'=\emptyset))=1$. That is, if $f=1$, then $A'$ is
empty. We say the flag $f_2$ of the call to
\textsc{Reorder}$(u,v,f_1,f_2)$ is correctly set only if
$f_2\Rightarrow(B'=\emptyset)=1$. That is, if $f_2=1$, then $B'$ is
empty.

\begin{lemma}\label{A:=0 iff A'=0}
Given a DAG $G$ with a valid topological order and two vertices $u,
v$ of $G$ with $u<v$, let $A' = \{w \mbox{ : } u\rightarrow w \mbox{
and } w< v \}$ and $B' = \{w \mbox{ : } w\rightarrow v \mbox{ and }
w> u \}$. If the flags are correctly set, then in the call of
\textsc{Reorder}$(u,v,f_1,f_2)$, $A:=\emptyset$ if and only if
$A'=\emptyset$. Similarly, if the flags are correctly set, then in
the call of \textsc{Reorder}$(u,v,f_1,f_2)$, $B:=\emptyset$ if and
only if $B'=\emptyset$.
\end{lemma}
\begin{proof}
We shall only prove that $A:=\emptyset$ if and only if
$A'=\emptyset$. It can be proved in a similar way that
$B:=\emptyset$ if and only if $B'=\emptyset$. There are two cases to
consider.

Case 1: $t_i<d(u,v)\leq t_{i+1}$ for some $i$ with $0<i\leq p$. By
the algorithm, $A:=\emptyset$ if and only if
    \[    \begin{array}{ll}
        \hat{A}_i=\{w \mbox{ : } u\rightarrow w \mbox{ and } d(u,w) \leq  t_i \mbox{ and } w < v\}=\emptyset \mbox{ and }\\
        (\hat{A}_{i+1} = \{w \mbox{ : } u\rightarrow w \mbox{ and } d(u,w) \leq t_{i+1} \mbox{ and } w
        < v\}=\emptyset \vee f_1=1).\\
        \end{array}
    \]
By $t_i<d(u,v)\leq t_{i+1}$, we have $\hat{A}_i\subseteq
\hat{A}_{i+1} = A'$. From $\hat{A}_i\subseteq \hat{A}_{i+1} = A'$
and $(f_1\Rightarrow (A'=\emptyset))=1$, we conclude that
$A:=\emptyset$ if and only if $A'=\emptyset$.

Case 2: $d(u,v)\leq t_0$. By the algorithm, $A:=\emptyset$ if and
only if
    \[    \begin{array}{ll}
        \hat{A}_0=\{w \mbox{ : } u\rightarrow w \mbox{ and } d(u,w) \leq  t_0 \mbox{ and } w < v\}=\emptyset.\\
        \end{array}
    \]
By $d(u,v)\leq t_0$, we have $\hat{A}_0=A'$. From $\hat{A}_0= A'$,
we conclude that $A:=\emptyset$ if and only if $A'=\emptyset$.
\end{proof}

\begin{lemma}\label{swap by itself}
Given a DAG $G$ with a valid topological order and two vertices $u,
v$ of $G$ with $u<v$, let $A' = \{w \mbox{ : } u\rightarrow w \mbox{
and } w< v \}$ and $B' = \{w \mbox{ : } w\rightarrow~v \mbox{ and }
w> u \}$. If the flags are correctly set, then
\textsc{Reorder}$(u,v,f_1,f_2)$ leads to a swap by itself if and
only if $A'=\emptyset$ and $B'=\emptyset$.
\end{lemma}
\begin{proof}
By the algorithm, the call to \textsc{Reorder}$(u,v,f_1,f_2)$ leads
to a swap by itself if and only if in this call $A:=\emptyset$ and
$B:=\emptyset$. By Lemma~\ref{A:=0 iff A'=0}, $A:=\emptyset$ and
$B:=\emptyset$ if and only if $A'=\emptyset$ and $B'=\emptyset$.
\end{proof}

Given a DAG $G$ with a valid topological order,
\textsc{Reorder}$(u,v,f_1,f_2)$ is said to be $local$ if and only if
the execution of \textsc{Reorder}$(u,v)$ will not affect $T(w)$ for
all $w$ with $w> v$ or $w<u$.

\begin{lemma}\label{stop and local}
Given a DAG $G$ with a valid topological order and two vertices
$u,v$ with $u\not\rightsquigarrow v$, if the flags are correctly
set, then \textsc{Reorder}$(u,v,f_1,f_2)$ maintains a valid
topological order and stop with $v< u$ and is local.
\end{lemma}
\begin{proof}
We prove the lemma by induction on $T(v)-T(u)$. When $T(u)-T(v)\leq
0$, the lemma is trivially correct.

Assume the lemma to be true when $T(v)-T(u)<k$, where $k>0$. We
shall prove that the lemma is true when $T(v)-T(u) = k$. If $A' =
\{w \mbox{ : } u\rightarrow w \mbox{ and } w< v \}=\emptyset$ and
$B' = \{w \mbox{ : } w\rightarrow~v \mbox{ and } w> u \}=\emptyset$,
then by Lemma~\ref{swap by itself}, line 13 is executed. Thus,
\textsc{Reorder}$(u,v,f_1,f_2)$ maintains a valid topological order,
stops with $v<u$, and only $T(u)$ and $T(v)$ are updated, so the
lemma follows. If $A' \neq\emptyset$ or $B' \neq \emptyset$, by
Lemma~\ref{swap by itself}, the \textbf{for}-loops are executed. Let
$T'$ be the initial topological order. By our induction hypothesis
and Lemma~\ref{A:=0 iff A'=0}, the following loop invariants hold:
 \[
        \begin{array}{ll}
        \mbox{1. $T$ is a valid topological order.}\\
        \mbox{2. At the start of the execution of line 19,
        $T(v')-T(u')<k$ and $T'(u)\leq T(u')< T(v') \leq T'(v)$.}\\
        \mbox{3. At the start of the execution of line 19,
        $u'\not\rightsquigarrow v'$.}\\
        \mbox{4. The flags are correctly set for the recursive call.}\\
        \end{array}
 \]
By the loop invariants and our induction hypothesis, each recursive
call \textsc{Reorder}$(u',v',f_1',f_2')$ in the \textbf{for}-loops
stops with $v'<u'$ and is local. Since the last recursive call is
\textsc{Reorder}$(u,v)$, the entire procedure stops with $v<u$.
Since each recursive call \textsc{Reorder}$(u',v')$ is local and
starts with $T'(u)\leq T(u')< T(v') \leq T'(v)$, the topological
order of vertices $w$ with $T'(w)>T'(v)$ or $T'(w)<T'(u)$ is not
affected. Thus the entire procedure maintains a valid topological
order, stops with $v<u$, and is local.
\end{proof}

\begin{theorem}\label{correctness}
Given a DAG $G$ with a valid topological order and two vertices $u,
v$ of $G$ with $u\not\rightsquigarrow v$, if the flags are correctly
set, then \textsc{Insert}$(u,v)$ will add an edge $(u,v)$ to $G$ and
maintain a valid topological order.
\end{theorem}
\begin{proof}
Because $u\not\rightsquigarrow v$, we know that $u$ and $v$ are two
different vertices and either $u < v$ or $u>v$. If $u < v$ then the
theorem is trivially correct. Assume that $v> u$. By Lemma~\ref{stop
and local}, \textsc{Reorder}$(v,u,0,0)$ will stop with $u< v$ and
maintain a valid topological order. Thus when line 2 of
\textsc{Insert} is ready to be executed, we will have a valid
topological order and $u<v$, and adding an edge $(u,v)$ to $G$ won't
affect the validness of the topological order.
\end{proof}

In addition to the correctness of the algorithm, we also want to
prove that the flags are always correctly set.

\begin{lemma}\label{correctly set flags for subsequent recursive calls}
Given a DAG $G$ with a valid topological order and two vertices $u,
v$ of $G$ with $u<v$, consider a call to
\textsc{Reorder}$(u,v,f_1,f_2)$. If the flags $f_1$ and $f_2$ are
correctly set, then while executing this call, all subsequent calls
to \textsc{Reorder} will also own correct flags.
\end{lemma}
\begin{proof}
We prove the lemma by induction on the depth of the recursion tree.
By Lemma~\ref{stop and local}, the call to
\textsc{Reorder}$(u,v,f_1,f_2)$ will stop, so the depth of the
recursion tree is finite. If the depth is zero, then no recursive
calls are made and the lemma follows.

Assume the lemma to be true when the depth of the recursion tree is
less than $k$, where $k>0$. We shall prove that the lemma is true
when the depth of the recursion tree is $k$. Since $k>0$, there is
at least one recursive call. Thus the \textbf{for}-loops are
executed. By Lemma~\ref{A:=0 iff A'=0} and Lemma~\ref{swap by
itself}, the following loop invariants hold:
 \[
        \begin{array}{ll}
        \mbox{1. $T$ is a valid topological order.}\\
        \mbox{2. At the start of the execution of line 19, $T(u')< T(v')$.}\\
        \mbox{3. At the start of the execution of line 19,
        $u'\not\rightsquigarrow v'$.}\\
        \mbox{4. The flags are correctly set for the recursive call.}\\
        \end{array}
 \]
By the loop invariants and our induction hypothesis, each recursive
call to \textsc{Reorder}$(u',v',f_1',f_2')$ in the
\textbf{for}-loops, together with all subsequent calls to
\textsc{Reorder} in it, own correct flags, and the lemma follows.
\end{proof}

\begin{theorem}\label{flags are always correctly set}
Given a DAG $G$ with a valid topological order and two vertices $u,
v$ of $G$ with $u\not\rightsquigarrow v$, while executing
\textsc{Insert}$(u,v)$, the flags are correctly set for all calls to
\textsc{Reorder}.
\end{theorem}
\begin{proof}
If $u<v$ then there are no calls to \textsc{Reorder} made while
executing \textsc{Insert}$(u,v)$, and the lemma follows. If $u>v$
then \textsc{Insert}$(u,v)$ will call \textsc{Reorder}$(v,u,0,0)$.
By Lemma~\ref{correctly set flags for subsequent recursive calls},
the call to \textsc{Reorder}$(v,u,0,0)$, together with all
subsequent calls to \textsc{Reorder} in it, own correct flags, and
the lemma follows.
\end{proof}

\section {Runtime}
In this section, we analyze the time required to insert a sequence
of edges. By Theorem~\ref{flags are always correctly set}, the flags
are always correctly set. To avoid unnecessary discussion, each
lemma, theorem, corollary, and proof in this section is state under
the assumption that the flags are correctly set. To avoid notational
overload, sometimes we shall just write \textsc{Reorder}$(u,v)$ and
ignore the flags.

\subsection{Properties}
\begin{lemma}\label{x,y,z}
Given a DAG $G$ with a valid topological order and two vertices $u$
and $v$ with $u<v$, then during the execution of
\textsc{Reorder}$(u,v)$, we have that (1) $T(x)$ is nondecreasing if
$u\rightsquigarrow x$; (2) $T(y)$ is nonincreasing if
$y\rightsquigarrow v$; and (3) $T(z)$ doesn't change if
$u\not\rightsquigarrow z$ and $z\not\rightsquigarrow v$.
\end{lemma}
\begin{proof}
We prove the lemma by induction on the depth of the recursion tree.
By Lemma~\ref{stop and local}, the call to \textsc{Reorder}$(u,v)$
will stop, so the depth of the recursion tree is finite. If the
depth is zero, then no recursive calls are made. It follows that
line 13 is executed, so the lemma follows.

Assume the lemma to be true when the depth of the recursion tree is
less than $k$, where $k>0$. We shall prove that the lemma is true
when the depth of the recursion tree is $k$. Since $k>0$, there is
at least one recursive call. Thus the \textbf{for}-loops are
executed. By Lemma~\ref{A:=0 iff A'=0} and Lemma~\ref{swap by
itself}, the following loop invariants hold:
 \[
        \begin{array}{ll}
        \mbox{1. $T$ is a valid topological order.}\\
        \mbox{2. At the start of the execution of line 19, $T(u')< T(v')$.}\\
        \mbox{3. At the start of the execution of line 19,
        $u'\not\rightsquigarrow v'$.}\\
        \mbox{4. The flags are correctly set for the recursive call.}\\
        \end{array}
 \]
Note that for each recursive call \textsc{Reorder}$(u',v')$ in the
\textbf{for}-loops, we have $u\rightsquigarrow u'$ and
$v'\rightsquigarrow v$. Let $u\rightsquigarrow x$. Then we have
either $u'\rightsquigarrow x$ or ($u'\not\rightsquigarrow x$ and
$x\not\rightsquigarrow v'$). By the induction hypothesis, $T(x)$ is
nondecreasing or doesn't change during the execution of the
recursive call. Thus $T(x)$ is nondecreasing if $u\rightsquigarrow
x$. Let $y\rightsquigarrow v$. Then we have either
$y\rightsquigarrow v'$ or ($u'\not\rightsquigarrow y$ and
$y\not\rightsquigarrow v'$). By the induction hypothesis, $T(y)$ is
nonincreasing or doesn't change during the execution of the
recursive call. Thus $T(y)$ is nonincreasing if $y\rightsquigarrow
v$. Let $u\not\rightsquigarrow z$ and $z\not\rightsquigarrow v$.
Then $u'\not\rightsquigarrow z$ and $z\not\rightsquigarrow v'$. By
the induction hypothesis, $T(z)$ doesn't change during the execution
of the recursive call. Thus $T(z)$ doesn't change if
$u\not\rightsquigarrow z$ and $z\not\rightsquigarrow v$.
\end{proof}

\begin{lemma}\label{at most one swap during a Reorder}
Given a DAG $G$ with a valid topological order and two vertices $u$
and $v$ with $u<v$, for all $x$ and $y$, \textsc{Reorder}$(u,v)$
leads to at most one swap of $x$ and $y$.
\end{lemma}
\begin{proof}
Suppose that \textsc{Reorder}$(u,v)$ leads to at least one swap of
$x$ and $y$. Without loss of generality we assume that $x<y$ before
the the first swap occurs. Then the first swap of $x$ and $y$ leads
to increase of $T(x)$ and decrease of $T(y)$. Thus by
Lemma~\ref{x,y,z}, $T(x)$ is nondecreasing and $T(y)$ is
nonincreasing during the execution of \textsc{Reorder}$(u,v)$. After
the first swap, we have $x>y$. Since $T(x)$ is nondecreasing and
$T(y)$ is nonincreasing, we know there are no more swaps.
\end{proof}

\begin{theorem}\label{keep the relative order after the first swap}
While inserting a sequence of edges, for all vertices $x$ and $y$,
after the first swap of $x$ and $y$, the relative order of $x$ and
$y$ won't change.
\end{theorem}
\begin{proof}
Suppose that the vertex pair $(x,y)$ is swapped at least once while
inserting a sequence of edges. By Lemma~\ref{at most one swap during
a Reorder} and the algorithm \textsc{Insert}, each edge insertion
leads to at most one swap of $x$ and $y$. Let $(u,v)$ be the first
edge whose insertion leads to a swap of $x$ and $y$. Without loss of
generality we assume that $x<y$ before the the first swap occurs. We
shall prove that $x>y$ will hold after the first swap of $x$ and
$y$. Consider the execution process of \textsc{Insert}$(u,v)$. The
first swap occurs during the execution of \textsc{Reorder}$(v,u)$.
By Lemma~\ref{x,y,z}, we have $v\rightsquigarrow x$ and
$y\rightsquigarrow u$. Since $T(x)$ is nondecreasing and $T(y)$ is
nonincreasing, after the first swap of $x$ and $y$, $x>y$ will hold
until \textsc{Reorder}$(v,u)$ returns. After \textsc{Reorder}$(v,u)$
returns, the edge $(u,v)$ will be added to the graph. Thus we will
have $y\rightsquigarrow x$ and $x>y$ just after
\textsc{Insert}$(u,v)$ returns. By Lemma~\ref{stop and local}, calls
to \textsc{Reorder} maintain a valid topological order, so $x>y$
will hold hereafter.
\end{proof}

\begin{corollary}\label{at most one swap}
While inserting a sequence of edges, for all vertices $x$ and $y$,
there is at most one swap of $x$ and $y$.
\end{corollary}

\begin{lemma}\label{at least one swap during a Reorder}
Given a DAG $G$ with a valid topological order and two vertices $u$
and $v$ with $u<v$, \textsc{Reorder}$(u,v)$ leads to a swap of $u$
and $v$.
\end{lemma}
\begin{proof}
We prove the lemma by induction on the depth of the recursion tree.
By Lemma~\ref{stop and local}, the call to \textsc{Reorder}$(u,v)$
will stop, so the depth of the recursion tree is finite. If the
depth is zero, then no recursive calls are made. It follows that
line 13 is executed, so the lemma follows.

Assume the lemma to be true when the depth of the recursion tree is
less than $k$, where $k>0$. We shall prove that the lemma is true
when the depth of the recursion tree is $k$. Since $k>0$, there is
at least one recursive call. Thus the \textbf{for}-loops are
executed. By Lemma~\ref{A:=0 iff A'=0} and Lemma~\ref{swap by
itself}, the following loop invariants hold:
 \[
        \begin{array}{ll}
        \mbox{1. $T$ is a valid topological order.}\\
        \mbox{2. At the start of the execution of line 19, $T(u')< T(v')$.}\\
        \mbox{3. At the start of the execution of line 19,
        $u'\not\rightsquigarrow v'$.}\\
        \mbox{4. The flags are correctly set for the recursive call.}\\
        \end{array}
 \]
Note that when executing the last recursive call
\textsc{Reorder}$(u',v')$ in the \textbf{for}-loops, we have $u'=u$
and $v'=v$. By our induction hypothesis and the loop invariants, the
last recursive call leads to a swap of $u$ and $v$, and the lemma
follows.
\end{proof}

\begin{theorem}\label{|A|+|B|}
While inserting a sequence of edges, the summation of $|A|+|B|$ over
all calls of \textsc{Reorder} is $O(n^2)$.
\end{theorem}
\begin{proof}
Consider arbitrary vertices $u$ and $\hat{v}$. We shall prove that
$\hat{v}\in B$ occurs at most once over all calls of
\textsc{Reorder}$(u,\cdot)$. This proves that the summation of $|B|$
over all calls of \textsc{Reorder}$(u,\cdot)$ is less than or equal
to $n$. Therefore the summation of $|B|$ over all calls of
\textsc{Reorder}$(\cdot,\cdot)$ is less than or equal to $n^2$.

Consider the execution process of the first call of
\textsc{Reorder}$(u,\cdot)$ for which $\hat{v}\in B$. By the
algorithm, a recursive call to \textsc{Reorder}$(u,\hat{v})$ is made
in the \textbf{for}-loops. Before the recursive call to
\textsc{Reorder}$(u,\hat{v})$ in the \textbf{for}-loops, at the
start of the execution of each recursive call to
\textsc{Reorder}$(u',v')$ in the \textbf{for}-loops, we have
$u<\hat{v}$ and ($u<u'<v'$ or $u = u' < v' <\hat{v}$). This follows
from the order in which we make the recursive calls and the local
property (Lemma~\ref{stop and local}). Since $u<\hat{v}$ and
($u<u'<v'$ or $u = u' < v' <\hat{v}$), by the local property,
$u<\hat{v}$ will hold during the execution of the call to
\textsc{Reorder}$(u',v')$. Thus before the recursive call to
\textsc{Reorder}$(u,\hat{v})$ in the \textbf{for}-loops, all
recursive calls to \textsc{Reorder}$(u',v')$ in the
\textbf{for}-loops won't lead to a call to
\textsc{Reorder}$(u,\cdot)$ for which $\hat{v}\in B$; otherwise, the
call to \textsc{Reorder}$(u,\cdot)$ for which $\hat{v}\in B$ will
lead to a call to \textsc{Reorder}$(u,\hat{v})$ which will further
lead to $\hat{v}>u$ by Lemma~\ref{stop and local}, leading to a
contradiction. Suppose for the contradiction that the recursive call
to \textsc{Reorder}$(u,\hat{v})$ in the \textbf{for}-loops leads to
a call to \textsc{Reorder}$(u,v'')$ for which $\hat{v}\in B$. By the
order in which we make the recursive calls and the local property,
at the start of the execution of the recursive call to
\textsc{Reorder}$(u,\hat{v})$ in the \textbf{for}-loops, we have
$u<\hat{v}$. Since $\hat{v}\rightsquigarrow v''$, at the start of
the execution of the recursive call to \textsc{Reorder}$(u,\hat{v})$
in the \textbf{for}-loops, we have $u<\hat{v}<v''$. Thus by the
local property, $v''>u$ will hold during the execution of the
recursive call to \textsc{Reorder}$(u,\hat{v})$ in the
\textbf{for}-loops. However, by Lemma~\ref{stop and local},
\textsc{Reorder}$(u,v'')$ stops with $v''<u$, which is a
contradiction. After the recursive call to
\textsc{Reorder}$(u,\hat{v})$ in the \textbf{for}-loops, we have
$\hat{v}<u$ by Lemma~\ref{stop and local}. Since $\hat{v}>u$ before
the recursive call to \textsc{Reorder}$(u,\hat{v})$ in the
\textbf{for}-loops, by Lemma~\ref{at least one swap during a
Reorder}, this recursive call leads to a swap of $u$ and $\hat{v}$.
Thus after the recursive call to \textsc{Reorder}$(u,\hat{v})$ in
the \textbf{for}-loops, we will have $\hat{v}<u$ and, by
Lemma~\ref{keep the relative order after the first swap}, the
relative order of $u$ and $\hat{v}$ won't change hereafter. Since
$\hat{v}<u$ holds hereafter, there will be no more calls of
\textsc{Reorder}$(u,\cdot)$ for which $\hat{v}\in B$. Putting all
things together, it follows that $\hat{v}\in B$ occurs at most once
over all calls of \textsc{Reorder}$(u,\cdot)$.

Similarly, we can prove that for arbitrary vertices $\hat{u}$ and
$v$, $\hat{u}\in A$ occurs at most once over all calls of
\textsc{Reorder}$(\cdot,v)$. It follows that the summation of $|A|$
over all calls of \textsc{Reorder}$(\cdot,\cdot)$ is less than or
equal to $n^2$.
\end{proof}

\begin{corollary}\label{O(n^2) times}
While inserting a sequence of edges, \textsc{Reorder} is called
$O(n^2)$ times.
\end{corollary}
\begin{proof}
By the algorithm, a call to \textsc{Reorder} for which $|A|+|B|=0$
leads to a swap by itself. By Corollary~\ref{at most one swap},
there are at most $n^2$ swaps. Thus there are at most $n^2$ calls to
\textsc{Reorder} for which $|A|+|B|=0$. By Theorem~\ref{|A|+|B|},
there are $O(n^2)$ calls to \textsc{Reorder} for which $|A|+|B|>0$.
Therefore, there are $O(n^2)$ calls to \textsc{Reorder} in total.
\end{proof}

Let $S=\{(u,v):$~ there is a swap of $u$ and $v$ such that $u<v$
while inserting the edges\}. Define
\[ D(u,v)=
    \left\{
        \begin{array}{ll}
             \mbox{the value of }d(u,v)\mbox{ while swapping $u$ and $v$}  & \forall(u,v)\in S;\\
             0 & \forall(u,v)\not\in S.
        \end{array}
    \right.
    \]

Since by Corollary~\ref{at most one swap} each vertex pair is
swapped at most once, $D(u,v)$ is well defined. Let $k$ be a number
with $1\leq k\leq n$. Define
\[ D_k(u,v)=
    \left\{
        \begin{array}{ll}
             D(u,v)  & \mbox{if } D(u,v)\leq k;\\
             k & \mbox{otherwise}.
        \end{array}
    \right.
    \]The following theorem is the key to our runtime
analysis.

\begin{theorem}\label{key}
For all $k\in[1,n]$ with  $k \geq n^{0.5}$, we have $\sum
D_k(u,v)=O(n^2\cdot\sqrt{k})$.
\end{theorem}
\begin{proof}
Let $k = n^{r}$. Let $T^*$ denote the final topological order.
Define $x(T^*(u),T^*(v))= D(u,v)$ and $z(T^*(u),T^*(v))= D_k(u,v)$
for all vertices $u$ and $v$. The following linear inequalities are
proved to be true by Ajwani~et~al. \cite{Ajwani}.

\[
        \begin{array}{ll}
        \mbox{(1) $x(i,j)\leq 0$ for all $1\leq i \leq n$ and $1\leq j \leq i$.}\\
        \mbox{(2) $x(i,j)\leq n$ for all $1\leq i \leq n$ and $i<j \leq n$.}\\
        \mbox{(3) $\sum_{j>i} x(i,j) - \sum_{j<i}x(j,i) \leq n$ for all $1\leq i \leq n$.}\\
        \end{array}
\]
It is easy to derive the following linear inequalities from the
definitions of $x(i,j)$ and $z(i,j)$.

\[
        \begin{array}{ll}
        \mbox{(4) $z(i,j)\leq n^r$ for all $1\leq i,j \leq n$.}\\
        \mbox{(5) $z(i,j) \leq x(i,j)$ for all $1\leq i,j \leq n$.}\\
        \mbox{(6) $0\leq z(i,j)$ for all $1\leq i \leq n$ and $1\leq j \leq n$.}\\
        \mbox{(7) $0\leq x(i,j)$ for all $1\leq i \leq n$ and $1\leq j \leq n$.}\\
        \end{array}
 \]
We aim to estimate an upper bound on the objective values of the
following linear program.

\[
        \begin{array}{ll}
        \mbox{max $\sum\limits_{1\leq i,j \leq n} z(i,j)$ such that}\\
        \end{array}
 \]
\[
        \begin{array}{ll}
        \mbox{(1) $x(i,j)\leq 0$ for all $1\leq i \leq n$ and $1\leq j \leq i$}\\
        \mbox{(2) $x(i,j)\leq n$ for all $1\leq i \leq n$ and $i<j \leq n$}\\
        \mbox{(3) $\sum_{j>i} x(i,j) - \sum_{j<i}x(j,i) \leq n$ for all $1\leq i \leq n$}\\
        \mbox{(4) $z(i,j)\leq n^r$ for all $1\leq i,j \leq n$}\\
        \mbox{(5) $z(i,j) \leq x(i,j)$ for all $1\leq i,j \leq n$}\\
        \mbox{(6) $0\leq z(i,j)$ for all $1\leq i \leq n$ and $1\leq j \leq n$}\\
        \mbox{(7) $0\leq x(i,j)$ for all $1\leq i \leq n$ and $1\leq j \leq n$.}\\
        \end{array}
 \]
In order to prove the upper bound on the objective values of the
above linear program, we consider its dual problem.
\[ \min  \left[
                    \begin{array}{ll}
                    \sum\limits_{0\leq i<j<n} n\cdot Y_{i\cdot n+j}+\sum\limits_{0\leq i<n}
                    n\cdot Y_{n^2+i}+\sum\limits_{0\leq i,j<n} n^r\cdot Z_{i\cdot n+j}
                    \end{array}
        \right]
        \mbox{ such that }
 \]
\[
        \begin{array}{ll}
        \mbox{(1) $Y_{i\cdot n+j}-W_{i\cdot n+j} \geq 0$ for all $0\leq i \leq n$ and $0\leq j \leq i$}\\
        \mbox{(2) $Y_{i\cdot n+j}-W_{i\cdot n +j}+Y_{n^2+i}-Y_{n^2+j}\geq 0$ for all $0\leq i < n$ and $j>i$}\\
        \mbox{(3) $Z_{i\cdot n+j}+W_{i\cdot n+j}\geq 1$ for all $0\leq i < n$ and $0\leq j<n$}\\
        \mbox{(4) $Y_i\geq 0$ for all $0\leq i < n^2+n$}\\
        \mbox{(5) $Z_i\geq 0$ for all $0\leq i < n^2$}\\
        \mbox{(6) $W_i\geq 0$ for all $0\leq i < n^2.$}\\
        \end{array}
 \]
Let $c$ be a large enough constant, e.g. 120, such that $(i+c\cdot
n^{r/2})^{r/2}\geq (i^{r/2}+1)$ for any $1\leq i\leq n$. The
following is a feasible solution to the dual problem.

\[ \left\{
        \begin{array}{ll}
        Y_{i\cdot n +j}=1 &\mbox{ for all } 0\leq i <n \mbox{ and } 0\leq j\leq i\\
        Z_{i\cdot n +j}=0 &\mbox{ for all } 0\leq i <n \mbox{ and } 0\leq j\leq i\\
        W_{i\cdot n +j}=1 &\mbox{ for all } 0\leq i <n \mbox{ and } 0\leq j\leq i\\
        Y_{i\cdot n +j}=0 &\mbox{ for all } 0\leq i <n \mbox{ and } i< j\leq i+c\cdot n^{r/2}\\
        Z_{i\cdot n +j}=1 &\mbox{ for all } 0\leq i <n \mbox{ and } i< j\leq i+c\cdot n^{r/2} \\
        W_{i\cdot n +j}=0 &\mbox{ for all } 0\leq i <n \mbox{ and } i< j\leq i+c\cdot n^{r/2} \\
        Y_{i\cdot n +j}=0 &\mbox{ for all } 0\leq i <n \mbox{ and } i+c\cdot n^{r/2}<j<n\\
        Z_{i\cdot n +j}=0 &\mbox{ for all } 0\leq i <n \mbox{ and } i+c\cdot n^{r/2}<j<n\\
        W_{i\cdot n +j}=1 &\mbox{ for all } 0\leq i <n \mbox{ and } i+c\cdot n^{r/2}<j<n\\
        Y_{n^2+i}=(n-i)^{r/2} &\mbox{ for all } 0\leq i <n\\
        \end{array}
    \right.
 \]
This feasible solution to the dual problem has an objective value of
$O(n\sum_{i=1}^ni^{r/2} + n\cdot n^r \cdot c\cdot
n^{r/2})=O(n^{2+r/2}+n^{1+r+r/2})=O(n^{2+r/2})=O(n^2\cdot
\sqrt{k})$, which by the primal-dual theorem is an upper bound on
the objective values of the original linear program.
\end{proof}

\begin{lemma}\label{A'=0 => f_1 =1 for subsequent calls to reorder(u,.)}
Given a DAG $G$ with a valid topological order and two vertices $u$
and $v$ with $u<v$, consider a call to \textsc{Reorder}$(u,v)$. If
$A' = \{w \mbox{ : } u\rightarrow w \mbox{ and } w< v \} =
\emptyset$, then when executing this call, we shall have the first
flag $f_1=1$ for all subsequent calls to
\textsc{Reorder}$(u,\cdot)$.
\end{lemma}
\begin{proof}
We prove the lemma by induction on the depth of the recursion tree.
By Lemma~\ref{stop and local}, the call to \textsc{Reorder}$(u,v)$
will stop, so the depth of the recursion tree is finite. If the
depth is zero, then no subsequent recursive calls are made, so the
lemma follows.

Assume the lemma to be true when the depth of the recursion tree is
less than $k$, where $k>0$. We shall prove that the lemma is true
when the depth of the recursion tree is $k$. Since $k>0$, there is
at least one recursive call. Thus the \textbf{for}-loops are
executed. By Lemma~\ref{A:=0 iff A'=0}, $A=\emptyset$. By the local
property (Lemma~\ref{swap by itself}) and the order in which we make
the recursive calls, any subsequent call to
\textsc{Reorder}$(u,\cdot)$ must occur during the execution of last
iteration of the outer \textbf{for}-loop. Consider any first level
recursive call \textsc{Reorder}$(u',v',f_1',f_2')$ in the last
iteration of the outer \textbf{for}-loop. Note that we have
$u'=u<v'$ and $\{w \mbox{ : } u\rightarrow w \mbox{ and } w< v'\leq
v \}\subseteq A' = \emptyset$ when this call is ready to be
executed. Since $u'=u$ and $A=\emptyset$, we also have $f_1'=1$. By
the induction hypothesis, we also have the first flag $f_1=1$ for
all subsequent calls to \textsc{Reorder}$(u,\cdot)$ while executing
this call. It completes the proof.
\end{proof}


\begin{lemma}\label{move u right to v}
Given a DAG $G$ with a valid topological order and two vertices $u$
and $v$ with $u<v$, let $A' = \{w \mbox{ : } u\rightarrow w \mbox{
and } w< v \} = \emptyset$. Then a call to \textsc{Reorder}$(u,v)$
will stop with $u$ at the initial position of $v$. That is, letting
$T_{before}$ be the topological order just before the call to
\textsc{Reorder}$(u,v)$, then the call to \textsc{Reorder}$(u,v)$
will return a topological order $T_{after}$ such that
$T_{after}(u)=T_{before}(v)$.
\end{lemma}
\begin{proof}
We prove the lemma by induction on the depth of the recursion tree.
By Lemma~\ref{stop and local}, the call to \textsc{Reorder}$(u,v)$
will stop, so the depth of the recursion tree is finite. If the
depth is zero, then no recursive calls are made. It follows that
line 13 is executed, so the lemma follows.

Assume the lemma to be true when the depth of the recursion tree is
less than $k$, where $k>0$. We shall prove that the lemma is true
when the depth of the recursion tree is $k$. Since $k>0$, there is
at least one recursive call. Thus the \textbf{for}-loops are
executed. Let $T_{before}$ be the initial topological order. By
Lemma~\ref{A:=0 iff A'=0}, Lemma~\ref{swap by itself}, and the
induction hypothesis, the following loop invariants hold:

 \[
        \begin{array}{ll}
        \mbox{1. $T$ is a valid topological order.}\\
        \mbox{2. At the start of the execution of line 19,
        $T(u)=T(u')<T(v')=T_{before}(v')$}.\\
        \mbox{3. At the start of the execution of line 19,
        $u'\not\rightsquigarrow v'$.}\\
        \mbox{4. After the execution of line 19, $T(u)=T_{before}(v')$}.\\
        \mbox{5. The flags are correctly set for the recursive call.}\\
        \end{array}
 \]
Note that for the last recursive call \textsc{Reorder}$(u',v')$ in
the \textbf{for}-loops, we have $v'= v$. Thus by the loop
invariants, we have $T(u)=T_{before}(v)$ after the last recursive
call in the \textbf{for}-loops.
\end{proof}


\begin{lemma}\label{move right t_i}
Given a DAG $G$ with a valid topological order and two vertices $u$
and $v$ with $u<v$, when executing a call of \textsc{Reorder}$(u,v)$
in which both $\hat{A}_i$ and $\hat{A}_{i+1}$ are computed, $u$ will
be moved right with distance at least $t_i$.
\end{lemma}
\begin{proof}
Since both $\hat{A}_i$ and $\hat{A}_{i+1}$ are computed, by the
algorithm, we have $t_i<d(u,v)\leq t_{i+1}$ and
$\hat{A}_i=\emptyset$. There are two cases to consider.

Case 1: $\hat{A}_{i+1}=\emptyset$. It follows that $A' = \{w \mbox{
: } u\rightarrow w \mbox{ and } w< v \}=\emptyset$. By
Lemma~\ref{move u right to v}, $u$ will be moved right to the
initial position of $v$. Since initially $d(u,v)> t_i$, $u$ will be
moved right with distance at least $t_i$.

Case 2: $\hat{A}_{i+1}\neq\emptyset$. By the algorithm, the
\textbf{for}-loops are executed. Let $\hat{u}$ be the vertex with
lowest topological order in $\hat{A}_{i+1}$. Let $T_{initial}$ be
the initial topological order. Since $\hat{A}_i=\emptyset$, we have
initially $d(u,\hat{u})> t_i$, i.e.,
$T_{initial}(\hat{u})-T_{initial}(v)>t_i$. By Lemma~\ref{stop and
local} and the order in which we make the recursive calls, before
the last iteration of the outer \textbf{for}-loop, $T(v)\geq
T_{initial}(\hat{u})$ will hold. Consider the execution of the last
iteration of the outer \textbf{for}-loop. Let $T_{start}$ be the
topological order at the start of this iteration. Then we have
$T_{start}(v)-T_{initial}(u)\geq
T_{initial}(\hat{u})-T_{initial}(u)>t_i$. By Lemma~\ref{stop and
local} and Lemma~\ref{move u right to v}, the following loop
invariants hold.
 \[
        \begin{array}{ll}
        \mbox{1. $T$ is a valid topological order.}\\
        \mbox{2. At the start of the execution of the last iteration of line 19,
        $T(u)=T(u')<T(v')=T_{start}(v')$.}\\
        \mbox{3. At the start of the execution of line 19,
        $u'\not\rightsquigarrow v'$.}\\
        \mbox{4. At the start of the execution of line 19, $\{w \mbox{ : } u\rightarrow w \mbox{ and } w<
        v'\leq v
        \}=\emptyset$}.\\
        \mbox{5. After the execution of line 19, $T(u)=T_{start}(v')$}.\\
        \mbox{6. The flags are correctly set for the recursive call.}\\
        \end{array}
 \]
Thus after this iteration, we will have
$T(u)=T_{start}(v)>T_{initial}(u)+t_i$, and the lemma follows.
\end{proof}


\begin{theorem}\label{both hatA_i and hatA_i+1 are computed}
While inserting a sequence of edges, there are at most $O(\frac{n^2
\sqrt{t_{i+1}}}{t_i})$ calls of \textsc{Reorder} for which both
$\hat{A}_{i}$ and $\hat{A}_{i+1}$ are computed. Similarly, there are
at most $O(\frac{n^2\sqrt{t_{i+1}}}{t_i})$ calls of \textsc{Reorder}
for which both $\hat{B}_{i}$ and $\hat{B}_{i+1}$ are computed.
\end{theorem}
\begin{proof}
We shall only prove that there are at most $O(\frac{n^2
\sqrt{t_{i+1}}}{t_i})$ calls of \textsc{Reorder} for which both
$\hat{A}_i$ and $\hat{A}_{i+1}$ are computed. It can be proved in a
similar way that there are at most
$O(\frac{n^2\sqrt{t_{i+1}}}{t_i})$ calls of \textsc{Reorder} for
which both $\hat{B}_{i}$ and $\hat{B}_{i+1}$ are computed.

Let $C_1(u,v),C_2(u,v),\ldots,C_{m(u,v)}(u,v)$ be the calls to
\textsc{Reorder}$(u,v)$ for which both $\hat{A}_i$ and
$\hat{A}_{i+1}$ are computed for all vertices $u$ and $v$. Let
$S_i(u,v)=\{w:$ $C_i(u,v)$ leads to a swap of $u$ and $w\}$ for
$i=1,\ldots,m(u,v)$. We shall prove that
$\sum_{u,v}\sum_{i=1}^{m(u,v)}\sum_{w\in S_i(u,v)} D(u,w)=
O(n^2\sqrt{t_{i+1}})$. By Lemma~\ref{move right t_i}, $\sum_{w\in
S_i(u,v)} D(u,w)\geq t_i$ for all vertices $u$ and $v$ and
$i=1,\ldots, m(u,v)$. It follows that $\sum_{u,v}\sum_{i=1}^{m(u,v)}
1= O(\frac{n^2\sqrt{t_{i+1}}}{t_i})$.

In a call to \textsc{Reorder}$(u,v)$, $\hat{A}_i$ and
$\hat{A}_{i+1}$ are both computed only if $t_i<d(u,v)\leq t_{i+1}.$
Thus by the local property (Lemma~\ref{stop and local}), we have
$D(u,w)\leq t_{i+1}$ for all $w\in S_i(u,v)$, $i=1,\ldots,m(u,v)$.
By Theorem~\ref{key}, we have
$\sum_{u,v}D_{t}(u,v)=O(n^2\sqrt{t_{i+1}})$. Thus it suffices to
show that in the summation $\sum_{u,v}\sum_{i=1}^{m(u,v)}\sum_{w\in
S_i(u,v)}D(u,w)$, $D(u,w)$ is counted at most twice for each vertex
pair $(u,w)$.

To show that in the summation
$\sum_{u,v}\sum_{i=1}^{m(u,v)}\sum_{w\in S_i(u,v)}D(u,w)$, $D(u,w)$
is counted at most twice for each vertex pair $(u,v)$, we only have
to prove that $S_i(u,v)\cap S_j(u,v')\cap S_k(u,v'')=\emptyset$ if
$C_i(u,v)$, $C_j(u,v')$, and $C_k(u,v'')$ are three different calls.

Suppose for the contradiction that $w\in S_i(u,v)\cap S_j(u,v')\cap
S_k(u,v'')$ and $C_i(u,v)$, $C_j(u,v')$, and $C_k(u,v'')$ are three
different calls. Without loss of generality, we assume $C_i(u,v)$
occurs before $C_j(u,v')$ and $C_j(u,v')$ occurs before
$C_k(u,v'')$. By Corollary~\ref{at most one swap}, there is only one
swap of $u$ and $w$, so $C_j(u,v')$ must be a subsequent recursive
call which occurs during the execution of $C_i(u,v)$ and
$C_k(u,v'')$ must be a subsequent recursive call which occurs during
the execution of $C_j(u,v')$. Consider the execution of $C_i(u,v)$.
By Lemma~\ref{stop and local} and the order in which we make the
recursive calls in the \textbf{for}-loops, $C_j(u,v')$ must occur
during the last iteration of the outer \textbf{for}-loop. Note that
by Lemma~\ref{stop and local}, before the last iteration of the
outer \textbf{for}-loop begins, all vertices in $\hat{A}_{i+1} = \{w
\mbox{ : } u\rightarrow w \mbox{, } d(u,w)\leq t_{i+1} \mbox{ and }
w < v\}=\{w \mbox{ : } u\rightarrow w \mbox{ and } w < v\}$ are
moved to the right of $v$. Thus when the last iteration of the outer
\textbf{for}-loop begins, there are not any vertices $w$ between $u$
and $v$ with $u\rightarrow w$. Therefore, by the local property of
\textsc{Reorder}, during the last iteration of the outer
\textbf{for}-loop, for each call to \textsc{Reorder}$(u,v')$, we
have $\{w: \mbox{$u\rightarrow w$ and $w<v'\leq v$}\}=\emptyset$. By
Lemma~\ref{A'=0 => f_1 =1 for subsequent calls to reorder(u,.)}, we
have the first flag $f_1=1$ for each subsequent call to
\textsc{Reorder}$(u,\cdot)$ during the execution of $C_j(u,v')$. It
follows that the first flag $f_1$ of the call $C_k(u,v'')$ is 1.
Thus by the algorithm, we don't compute $\hat{A}_{i+1}$ in the call
$C_k(u,v'')$, which is a contradiction.
\end{proof}

\subsection{Runtime Analysis}

\begin{lemma}\label{edge insertion}
While inserting a sequence of edges, the total time spent on
executing line 2 of \textsc{Insert} is $\tilde{O}(n^2)$.
\end{lemma}
\begin{proof}
As discussed in Section~\ref{Instructions}, each execution of line 2
of \textsc{Insert} can be done in $\tilde{O}(1)$ time. Since there
are at most $n(n-1)/2$ edge insertions, the lemma follows.
\end{proof}

\begin{lemma}\label{compute hatA_i and hatB_i when t_i<d(u,v)<=t_i+1}
While inserting a sequence of edges, the total time spent on
computing $\hat{A}_i$ and $\hat{B}_i$, $i=1,\ldots,p$, over all
calls of \textsc{Reorder}$(u,v)$ with $t_i<d(u,v)\leq t_{i+1}$ is
$\tilde{O}(n^2)$.
\end{lemma}
\begin{proof}
As discussed in Section~\ref{Instructions}, it needs
$\tilde{O}(|\hat{A}_i|+|\hat{B}_i|+1)$ time to compute $\hat{A}_i$
and $\hat{B}_i$ in a call of \textsc{Reorder}$(u,v)$ if
$d(u,v)<t_{i+1}$. By Theorem~\ref{|A|+|B|}, the summation of
$|\hat{A}_i|+|\hat{B}_i|$, $i=1,\ldots,p$, over all calls of
\textsc{Reorder} is $O(n^2)$. By Corollary~\ref{O(n^2) times},
\textsc{Reorder} is called $O(n^2)$ times. Thus the summation of
$(|\hat{A}_i|+|\hat{B}_i|+1)$ over all calls of \textsc{Reorder} is
$O(n^2)$, and the lemma follows.
\end{proof}

\begin{lemma}\label{compute hatA_i and hatB_i when d(u,v)<=t_i}
For each $i$ with $1\leq i\leq p+1$, while inserting a sequence of
edges, the total time spent on computing $\hat{A}_i$ and
$\hat{B}_i$, over all calls of \textsc{Reorder}$(u,v)$ with
$d(u,v)\leq t_{i}$ is $O(\frac{n^2t_i^{3/2}}{t_{i-1}})$.
\end{lemma}
\begin{proof}
If $d(u,v)\leq t_{i}$, then by the algorithm, $\hat{A}_i$ is
computed only if $\hat{A}_{i-1}$ is also computed and is empty. Thus
by Theorem~\ref{both hatA_i and hatA_i+1 are computed}, there are at
most $O(\frac{n^2t_i^{1/2}}{t_{i-1}})$ such calls. As discussed in
Section~\ref{Instructions}, it needs $O(t_i)$ time to compute
$\hat{A}_i$ in each of such calls. Thus the total time spent on
computing $\hat{A}_i$ over all calls of \textsc{Reorder}$(u,v)$ with
$d(u,v)\leq t_{i}$ is $O(\frac{n^2t_i^{3/2}}{t_{i-1}})$. Similarly,
the total time spent on computing $\hat{B}_i$ over all calls of
\textsc{Reorder}$(u,v)$ with $d(u,v)\leq t_{i}$ is
$O(\frac{n^2t_i^{3/2}}{t_{i-1}})$.
\end{proof}

\begin{lemma}\label{compute hatA_0 and hatB_0}
While inserting a sequence of edges, the total time spent on
computing $\hat{A}_0$ and $\hat{B}_0$ over all calls of
\textsc{Reorder} is $O(n^2\cdot t_0)$.
\end{lemma}
\begin{proof}
As discussed in Section~\ref{Instructions}, it needs $O(t_0)$ time
to compute $\hat{A}_0$ and $\hat{B}_0$ in a call of
\textsc{Reorder}. By Corollary~\ref{O(n^2) times}, \textsc{Reorder}
is called $O(n^2)$ times, and the lemma follows.
\end{proof}

\begin{lemma}\label{swap u and v}
While inserting a sequence of edges, the total time spent on
swapping vertices is $\tilde{O}(n^{2.5})$.
\end{lemma}
\begin{proof}
As discussed in Section~\ref{Instructions}, each swap of vertices
$u$ and $v$ with $d(u,v)<t$ can be done in $\tilde{O}(d(u,v))$ time.
By Theorem~\ref{key}, $\sum D_n(u,v)=O(n^{2.5})$. Thus the total
time is $\tilde{O}(n^{2.5})$.
\end{proof}

\begin{theorem}\label{total time}
While inserting a sequence of edges, the total time required is
$\tilde{O}(\sum_{i=1}^{p+1}\frac{n^2t_{i}^{3/2}}{ t_{i-1}}+n^2\cdot
t_0)$.
\end{theorem}
\begin{proof}
It follows directly from the above lemmas.
\end{proof}

\section {Further Discussion} \label{discussion}
Let $n^{0.5}<t_0<t_1<\ldots<t_p<t_{p+1}=n$ and $p=O(\log n)$. We
have known that the runtime is
$\tilde{O}(\sum_{i=1}^{p+1}\frac{n^2t_{i}^{3/2}}{ t_{i-1}}+n^2\cdot
t_0)$. In this section, we show how to determine the values of these
parameters. By letting
\[\frac{n^3}{\sqrt{t_{p+1}}}=\frac{n^2t_{p+1}^{3/2}}{ t_{p}}
=\frac{n^2t_{p}^{3/2}}{ t_{p-1}}=\frac{n^2t_{p-1}^{3/2}}{
t_{p-2}},\ldots,=\frac{n^2t_{1}^{3/2}}{ t_{0}}=n^2\cdot t_0,\] we
have
\[t_1=t_0^{4/3}, t_i=\frac{t_{i-1}^{5/3}}{t_{i-2}^{2/3}} \mbox{ for all }i = 2,\ldots, p+1.\] Let $t_i=t_0^{x_i}$ for $i=0,\ldots,p+1$. we have
\[x_0=1,x_1=4/3, \mbox{ and } x_i=\frac{5x_{i-1}}{3}-\frac{2x_{i-2}}{3} \mbox{ for all } i = 2,\ldots, p+1
.\] By solving this linear second order recurrence relation, we get
$x_i=2-(2/3)^i$ for all $i=0,1,2,\ldots,p+1$. It follows that
$t_{p+1}=t_0^{2-(2/3)^{(p+1)}}$. Since $t_{p+1}=n$, we have $t_0=
n^{f(p)}$, where $f(p)=\frac{3^{p+1}}{(2\cdot 3^{p+1}-2^{p+1})}.$

\begin{corollary}\label{general p}
While inserting a sequence of edges, the total time required is
$\tilde{O}(n^{2+f(p)})$ if $t_0=n^{f(p)}$ and $t_i=t_0^{2-(2/3)^i}$
for all $i=1,\ldots,p+1$, where $f(p)=\frac{3^{p+1}}{(2\cdot
3^{p+1}-2^{p+1})}.$
\end{corollary}

Note that $f(p)=\frac{3^{p+1}}{(2\cdot
3^{p+1}-2^{p+1})}=0.5+\frac{2^{p}}{(2\cdot 3^{p+1}-2^{p+1})}$. By
letting $\epsilon(p)=\frac{2^p}{(2\cdot 3^{p+1}-2^{p+1})}$, we have
$\epsilon(p)<\frac{2^p}{3^{p+1}} < (3/2)^{-p}$. By letting
$p=\log_{3/2} n$, we have $1<n^{\epsilon(p)}<n^{1/n}<2$ when $n>2$.
Thus
$\tilde{O}(n^{2+f(p)})=\tilde{O}(n^{2.5+\epsilon(p)})=\tilde{O}(n^{2.5})$
if we choose $p=\lceil\log_{3/2} n\rceil$. The following theorem
summarizes our discussion.

\begin{theorem}\label{result}
There exists an $\tilde{O}(n^{2.5})$-time algorithm for online
topological ordering.
\end{theorem}


\section {Concluding Remarks}
We propose an $\tilde{O}(n^{2.5})$-time algorithm for maintaining
the topological order of a DAG with $n$ vertices while inserting $m$
edges. By combining this with the result in \cite{Katriel 2006}, we
get an upper bound of $\tilde{O}(\min\{m^{3/2}, n^{2.5}\})$ for
online topological ordering. The only non-trivial lower bound is due
to Ramalingam and Reps \cite{Ramalingam}, who show that any
algorithm need $\Omega(n\log n)$ time while inserting $n-1$ edges in
the worst case if all labels are maintained explicitly. Precisely,
our algorithm runs in $O(n^{2.5}\log^2 n)$ time. Choosing a better
implementation for the pails, like data structures discussed in
Section~5 of~\cite{Ajwani}, can further drop one log factor from the
runtime.

\section*{Acknowledgments}
We thank Cheng-Wei Luo for helpful discussion.

\end{document}